\documentclass[twocolumn,tighten,times]{aastex63}

\hypersetup{colorlinks=true,linkcolor=blue,citecolor=blue,urlcolor=blue,}
\usepackage{verbatim} 
\usepackage{color,soul} 

\usepackage{lineno}

\shorttitle{\sc Minor Axis {{\CIV}} Absorption}
\shortauthors{\sc Croom \etal}

\input{iondefs.sty}

\begin{document}

\title{Circumgalactic C\,{\small\bf IV} Absorption Found Only Within {\boldmath $30^\circ$} of the Major Axes of Present-Epoch Galaxies}

\author[0000-0002-2908-9702]{Mark A. Croom}
\affiliation{Department of Astronomy, New Mexico State University, Las Cruces, NM 88003, USA}

\author[0000-0002-9125-8159]{Christopher W. Churchill}
\affiliation{Department of Astronomy, New Mexico State University, Las Cruces, NM 88003, USA}

\author[0000-0002-0507-7096]{Bart Wakker}
\affiliation{Eureka Scientific, Oakland, CA 94602, USA}

\author[0000-0003-1362-9302]{Glenn G. Kacprzak}
\affiliation{Centre for Astrophysics and Supercomputing, Swinburne University of Technology, Hawthorn, Victoria 3122, Australia}

\author[0000-0003-2377-8352]{Nikole M. Nielsen}
\affiliation{Homer L. Dodge Department of Physics and Astronomy, The University of Oklahoma, 440 W. Brooks St., Norman, OK 73019, USA}
\affiliation{Centre for Astrophysics and Supercomputing, Swinburne University of Technology, Hawthorn, Victoria 3122, Australia}

\begin{abstract}
We studied {\CIVdblt} absorption in the circumgalactic medium (CGM) of 88 isolated, bright ($L_B \geq 0.2L^*_B$), galaxies at $z\leq 0.03$ selected without {\it a priori\/} knowledge of CGM absorption. The galaxies have observationally-unbiased distributions of inclination and galaxy--quasar sight line azimuthal angle, $\Phi$.  Above a $5\sigma$ rest-frame equivalent width detection threshold of $W_r \geq 0.1$~{\AA}, we found absorption in the CGM of 9/88 of the galaxies within projected separation $R_\perp \leq 350$~kpc. All nine absorbers arise within $\Delta \Phi  = 30^\circ$ of the projected minor axes and $R_\perp \leq 1.2R_{\rm vir}$ of star-forming galaxies ($\log ({\rm sSFR}/{\rm yr}^{-1}) \geq -10.75$); the probability that this result is random chance is $P = 2 \times 10^{-5}$, corresponding to a 99.998\% ($4.3\sigma$) confidence level in our finding.
The data suggest strong evidence for biconical polar winds related to the moderately elevated star formation in these present-epoch galaxies and we describe simple spatial-kinematic models that support this hypothesis. High specific star-formation rates may not be a sufficient condition for predicting minor-axis {\CIV} absorption because the covering fraction of the subsample of star-forming galaxies within $30^\circ$ of their minor axes  is $f_c \sim 0.5$; this may imply transient and/or patchy outflows. Incorporating studies of {\MgII} and {\OVI} CGM absorption, we favor a scenario in which {\CIV}-bearing CGM gas interfaces between these low- and high-ionization regimes, comprising enriched intermediate-ionization clouds entrained and collimated in biconical stellar winds while mixing and cooling via non-equilibrium photoionization.
 
\end{abstract}


\section{Introduction}
\label{sec:intro}

The circumgalactic medium (CGM) is thought to regulate galaxy growth through the balancing of key processes such as gas accretion, stellar-feedback driven outflows, recycling, and mixing \citep{tumlinson17, peroux20, faucher-giguere23}. Simulations generally predict a geometrically-structured CGM in which cool, low-metallicity accretion preferentially aligns with the disk plane, while metal-enriched biconical outflows populate the polar axis region \citep[e.g.,][]{peroux20, defilippis21, hafen22, trapp22, stern24, yang24}. Observationally characterizing any spatial-kinematic trends using quasar absorption lines studies can potentially provide insights into the global processes of CGM baryon cycles \citep[e.g.,][]{skamI, skamII}.

For the most part, observations show that low-ionization gas traced by {\MgII} arises primarily along both the projected major and minor axes of their host galaxies \citep{bordoloi11, bouche12, kacprzak12, lan14}. 
Spatial-kinematic observations are consistent with simple models in which major-axis {\MgII} absorption is co-rotating and/or spiraling inward, whereas minor-axis absorption is entrained in outflowing biconical stellar-driven winds \citep[e.g.,][]{steidel02, kacprzak11kin, magiicat5, bouche16, ho17, rahmani18, martin19,  schroetter19, zabl19, zabl20,lopez20}. 

Higher-ionization gas provides a different, yet complementary view of the CGM. Observationally, stronger {\OVI} absorbers are found primarily around star-forming galaxies \citep{tumlinson11, tchernyshyov22} and along both their projected major and minor axes \citep{kacprzak15, beckett21, dutta26}. However, the interpretation of {\OVI} absorption is less clear than for {\MgII} as {\OVI} kinematics appear to not be coupled to host-galaxy rotation or indicative of polar outflows \citep{nielsen17, kacprzak19kine, kacprzak25, nateghi24GFII, ho26}. 

The intermediate-ionization gas phase probed by {\CIV} absorption is also common in the CGM of galaxies \citep{chen01a, bordoloi-cosdwarfs}. The absorption strengths and covering fractions tend to be larger for higher stellar mass and star-forming, post-burst, and/or actively starburst galaxies \citep{borthakur13, burchett16, hernandez22, garza25a}. These studies suggest that {\CIV} absorption may provide important insights for linking warm entrained material to the hotter biconical/polar outflowing medium \citep[e.g.,][]{fielding20}.

Unlike {\MgII} and {\OVI} CGM absorption, which have been shown to prefer the planar and polar azimuthal regions of their host galaxies, the azimuthal dependence of {\CIV} is not observationally constrained. 
In this {\it Letter}, we report on the absorption strengths and covering fractions of {\CIV} absorption as a function of $R_\perp/R_{\rm vir}$ and azimuthal-angle, $\Phi$, for a sample of 88 present-epoch galaxies with $L_B\geq 0.2 L^*_B $. In Section~\ref{sec:data}, we describe the data catalogs, the galaxy selection criteria, and the absorption line analysis.  In Section~\ref{sec:theCIVCGM}, we present our findings and in Section~\ref{sec:wrapitup} we summarize our findings. Throughout, we adopted the ``737 Cosmology" parameters  ${h = H_0/100 = 0.7}$, ${\Omega_m = 0.3}$,  and ${\Omega_\Lambda = 0.7}$.

\section{The Data and Sample}
\label{sec:data}

We built a catalog\footnote{We used a version dated August 2025.} of galaxies in fields toward quasar sightlines for which COS spectra could be retrieved from MAST. The galaxy catalog was created by retrieving NED parameters for 130,000 galaxies with spectroscopic redshifts \citep[Wakker et~al.\ in preparation; also see][]{french17}. This database was then used to derive regularized and homogenized galaxy parameters, including luminosity and redshift-independent distance (if available). Otherwise, distances were derived from the redshift corrected for Virgocentric flow \citep[e.g.,][]{huchra_groups_1982}. 


\begin{deluxetable*}{lccrcccc}
\tablecaption{Galaxy Sample\label{tab:galaxies}}
\tablewidth{0pt}
\tablehead{
\colhead{Galaxy} & 
\colhead{$z_{\rm gal}$} & 
\colhead{$\log(\rm sSFR)$\tablenotemark{a}} & 
\colhead{QSO} & 
\colhead{$R_{\perp}$} & 
\colhead{$R_{\perp}/R_{\rm vir}$\tablenotemark{b}} & 
\colhead{$\Phi$\tablenotemark{c}} & 
\colhead{$W_r(1548)$}  \\[-6pt]
\colhead{} & 
\colhead{} & 
\colhead{[$\log(\rm yr^{-1})$]} & 
\colhead{} & 
\colhead{[kpc]} & 
\colhead{} & 
\colhead{[deg]} & 
\colhead{[\AA]} 
}
\startdata
NGC 0755  & 0.00546 & $-10.204$   & J015530.02$-$085704.2 &  83 & 0.40 & 65 & 0.222 \\[-4pt]
NGC 1097  & 0.00424 & $-10.243$  & J024649.87$-$300741.5 &  51 & 0.12 & 83 & 0.338 \\[-4pt]
NGC 2611  & 0.01752 & $-9.840$   & J083535.81$+$245940.2 &  54 & 0.23 & 89 & 1.237 \\[-4pt]
NGC 3067  & 0.00492 & $-9.979$   & J095820.95$+$322402.2 &  10 & 0.05 & 75 & 1.345 \\[-4pt]
NGC 3173  & 0.00840 & $-10.585$  & J101558.90$-$274827.0 & 218 & 0.85 & 89 & 0.344 \\[-4pt]
NGC 3489  & 0.00226 & \nodata  & J105945.24$+$144143.0 & 106 & 0.53 & 80 & 0.254 \\[-4pt]
NGC 3941  & 0.00310 & \nodata  & J114646.03$+$371511.8 & 262 & 1.11 & 87 & 0.123 \\[-4pt]
NGC 4253  & 0.01292 & $-9.882$   & J121752.08$+$300700.6 & 331 & 1.18 & 85 & 0.132 \\[-4pt]
NGC 4517  & 0.00376 & $-10.533$  & J123304.05$-$003134.1 & 108 & 0.48 & 90 & 0.128 \\
\enddata
\tablecomments{Table 1 is published in its entirety in the machine-readable format. ($a$) As adopted from SGA2020 \citep{Moustakas_2023}. ($b$) As measured by \citet{french17}. ($c$) computed from the galaxy sky PAs adopted from HECATE \citep{2023hell.confE...1K}.}
\end{deluxetable*}

The observed galaxy distances and {\it BVR\/} and $ugr$ magnitudes were used to compute absolute magnitudes. Scaling to solar luminosity \citep{willmer18}, and scaling to $M^*_B$ \citep{marzke_luminosity_1994, loveday12} yielded the  $L_B/L^*_B$ luminosities. This is converted to a stellar mass, $M_*$, using the relations between $M/L$ and $g$--$r$ or $B$--$V$ color \citep{bell03}. Stellar mass is then converted into a halo mass, $M_h$, using the relation in \citet{moster13}. The virial radius, $R_{\rm vir}$, was determined for mean overdensity $\Delta_c=200$.

The star formation rates, galaxy inclinations, and galaxy--quasar azimuthal angles were drawn from The Heraklion Extragalactic Catalog \citep[HECATE,][]{2023hell.confE...1K}, which provides homogenized properties of 204,733 $cz \leq 14,000$~{\kms} galaxies. HECATE orientations were compared with the Siena Galaxy Atlas 2020 \citep{Moustakas_2023}, which served as the source in a few cases. 

\subsection{Selection of COS Spectra}

We then retrieved the COS spectra. Fixed-pattern noise features not corrected by CalCOS were filtered out. 
We then produced a single combined G160M spectrum for each quasar using the methods of \citet{wakker_nearby_2015}.
To build our galaxy sample, we begin by selecting quasar sightlines for which the average signal-to-noise ratio is  ${{\rm SNR} \geq 7}$ in the COS/G160M spectrum  over the wavelength range 1550--1598~{\AA}, corresponding to {\CIV} absorption at ${0.0018 \leq z \leq 0.03}$.  We limited our search to $cz > 500$~{\kms} to avoid Local Group galaxies.  The galaxy catalog is nearly 100\% complete to $L_B/L^*_B \geq 0.1$ out to $cz=10,000$~{\kms} ($z \leq 0.03$). We thus also limited our search to $z \leq 0.03$ in order to enforce a uniform luminosity completeness. This yielded 356 quasar sightlines.

\subsection{Selection of Intervening Galaxies}

We then searched these quasar sightlines for bright ($L_B\geq 0.2 L^*_B$) galaxies with projected separations of $R_\perp/R_{\rm vir} \leq 1.5$ from the quasar line of sight, where $R_\perp$ is the galaxy--quasar impact parameter and $R_{\rm vir}$ is the galaxy virial radius. We adopt this minimum luminosity to exclude dwarf galaxies from the COS-Dwarfs study \citep{bordoloi-cosdwarfs}. We compiled this initial list of galaxies blind to absorption in the COS spectra and to all other galaxy properties. 

We then examine the environment of each selected galaxy. We classified a galaxy as an isolated potential absorbing galaxy if the 3D distance between a potential neighbor galaxy was greater than two times the sum of the $R_{\rm vir}$ of the candidate and the potential neighbor galaxy. We also enforced a line-of-sight velocity separation of $\Delta v \geq \pm 400$~{\kms}. This velocity was adopted based on observed distribution of the two-point velocity correlation function (TPCF) of {\CIV}-selected absorbers, which shows no power for velocity separations greater than 400~{\kms} \citep{rauch96, boksy15}. 

These selection criteria yielded a sample of 88 isolated potential absorbing galaxies. The selected galaxies have no LMC-like or brighter satellites with $L_B/L^*_B \geq 0.1$. The redshift range of the galaxies is ${0.018 \leq z \leq  0.033}$ and the range of luminosities, stellar masses, and halo masses are ${0.2 \leq L_B/L^*_B \leq  8.2}$, ${8.1 \leq \log (M_*/M_{\odot}) \leq 11.4}$, and  ${11.8 \leq \log (M_{\rm h}/M_{\odot}) \leq 13.2}$. The range of inclinations is ${16.9^{\circ} \leq i \leq  90^\circ}$ and the range of galaxy--quasar azimuthal angles is ${1.5^{\circ} \leq \Phi \leq  89.9^\circ}$. We conducted one-sample KS tests comparing the inclination and azimuthal angle distributions to their respective unbiased distributions. For galaxy inclination, we obtained ${\rm KS} = 0.110$ with $P({\rm KS}) = 2.1\times 10^{-1}$ ($\sigma_{\hbox{\tiny KS}} = 0.17$), and for $\Phi$, we obtained ${\rm KS} = 0.122$ with $P({\rm KS}) = 1.3\times 10^{-1}$ ($\sigma_{\hbox{\tiny KS}} = 0.28$). The observed distributions of galaxy orientations are highly consistent with observationally unbiased distributions.








\subsection{Finding Associated Absorption}

After we built our isolated galaxy sample, we undertook an independent blind search for {\CIV} absorption. We used the tool \textsc{Search} \citep[e.g.][]{cwc-thesis, weakI}, which uses the algorithms described by \citet{schneider93}, to objectively locate and identify {\CIV} doublets in COS/G160M spectral segments spanning a velocity window ${\Delta v = \pm 1000}$~{\kms} around each galaxy redshift. The continuum of each spectral segment was first interactively continuum fitted using  \textsc{Linetools} \citep{prochaska16linetools}. 

We required a $5\sigma$ significance level for the $\lambda 1548$ transition and a $3\sigma$ significance level for the $\lambda 1550$ transition. 
In rare cases we needed to adjudicate complexity or ambiguity in the {\CIVdblt} identifications to rule out the presence of potential {\Lya} or {\Lyb} forest lines  masquerading as a {\CIV} line. In these cases, we consulted a semi-automated and visually vetted absorption line database we built for each quasar spectrum. In the cases where no {\CIV} absorption was identified, we quote the $3\sigma$ upper limit on the rest-frame equivalent width at the redshift of the potential absorbing galaxy. The rest-frame equivalent width detection threshold is 98\% complete to $W_r(\lambda 1548) = 0.1$~{\AA} at the $5\sigma$ significance level.

\section{Results}
\label{sec:theCIVCGM}

Of the 88 galaxies in the sample, nine were found to have {\CIV} CGM absorption.
The rest-frame {\CIV} equivalent width range is ${0.123 \leq W_r \leq 1.345}$~{\AA}, and the galaxy--absorber velocity offset range is ${-105 \leq \Delta v_{\rm g/a} \leq +30}$~{\kms}. For the remaining 79 galaxies, the $3\sigma$ rest-frame equivalent width limits ranged from ${0.006 \leq W_{r,\rm lim} \leq 0.092}$~{\AA}.

In Table~\ref{tab:galaxies}, we list the quantities discussed in this work; the printed snippet shows the nine galaxies with detected {\CIV} CGM absorption. These include the galaxy specific star formation rate, the physical impact parameter and the virial-radius-normalized impact parameter, the galaxy-quasar azimuthal angle, and the measured {\CIV}~$\lambda 1548$ rest-frame equivalent width. The complete sample is available electronically; the tabulated equivalent widths of the 79 non-absorbing cases, with ${W_r < 0.1}$~{\AA}, are their $3\sigma$ rest-frame equivalent width limits.

\subsection{Stellar Mass and Star Formation}
\label{sec:coulditbesSFR}

\begin{figure}[h!bt]
\centering
\includegraphics[width=1.0\columnwidth]{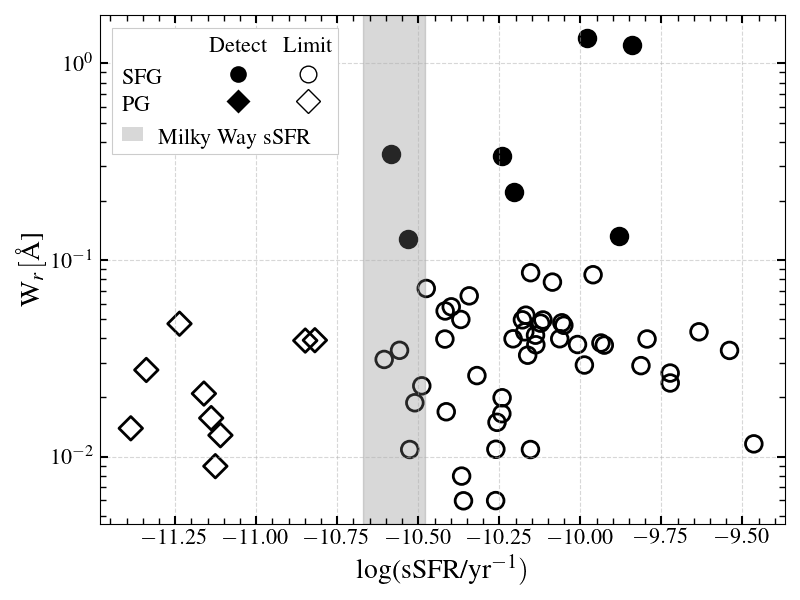}
\caption{
The  {\CIV} $\lambda 1548$ equivalent width, $W_r$, versus the specific star-formation rate (sSFR) for sample galaxies with measured stellar masses and star formation rates. Circle points are the subsample of star-forming galaxies (SFGs), diamond points are passive galaxies (PGs). Filled points have detected absorption and open points are upper limits. The gray shaded area is the measured range of sSFR for the Milky Way from \citet{licquia15}.}
\label{fig:EWvssSFR}
\end{figure}

In Figure~\ref{fig:EWvssSFR}, we show $W_r$ versus specific star formation rate (sSFR) for the subsample of galaxies with measured stellar masses and star formation rates. The measured range for the Milky Way is shown as the gray shaded region. With an sSFR of ${-10.67 \leq \log ({\rm sSFR}/{\rm yr}^{-1}) \leq -10.48}$ \citep[][]{licquia15}, the Milky Way is somewhat low star-forming galaxy, yet it lies within $2\sigma$ of the mean of the local-galaxy star-forming main sequence \citep{fraser-mckelvie19}.


Assuming the Milky Way galaxy is a lower boundary star-forming galaxy, we adopt ${\log ({\rm sSFR}/{\rm yr}^{-1}) = -10.75}$ to bifurcate the galaxy sample between lower sSFR passive galaxies (PGs) and higher sSFR star-forming galaxies (SFGs). For the seven sample galaxies with measured sSFR, those with detected {\CIV} absorbing gas are all SFGs, whereas none of the PGs exhibit detectable {\CIV} absorption. However, the majority of the SFGs also lack detected absorption. We thus conclude that, although {\CIV} absorbing gas may be found around SFGs only, high sSFR is not a sufficient condition for predicting {\CIV} absorption with $W_r \geq 0.1$~{\AA}.

\begin{figure*}[b!th]
\centering
\includegraphics[width=2.1\columnwidth]{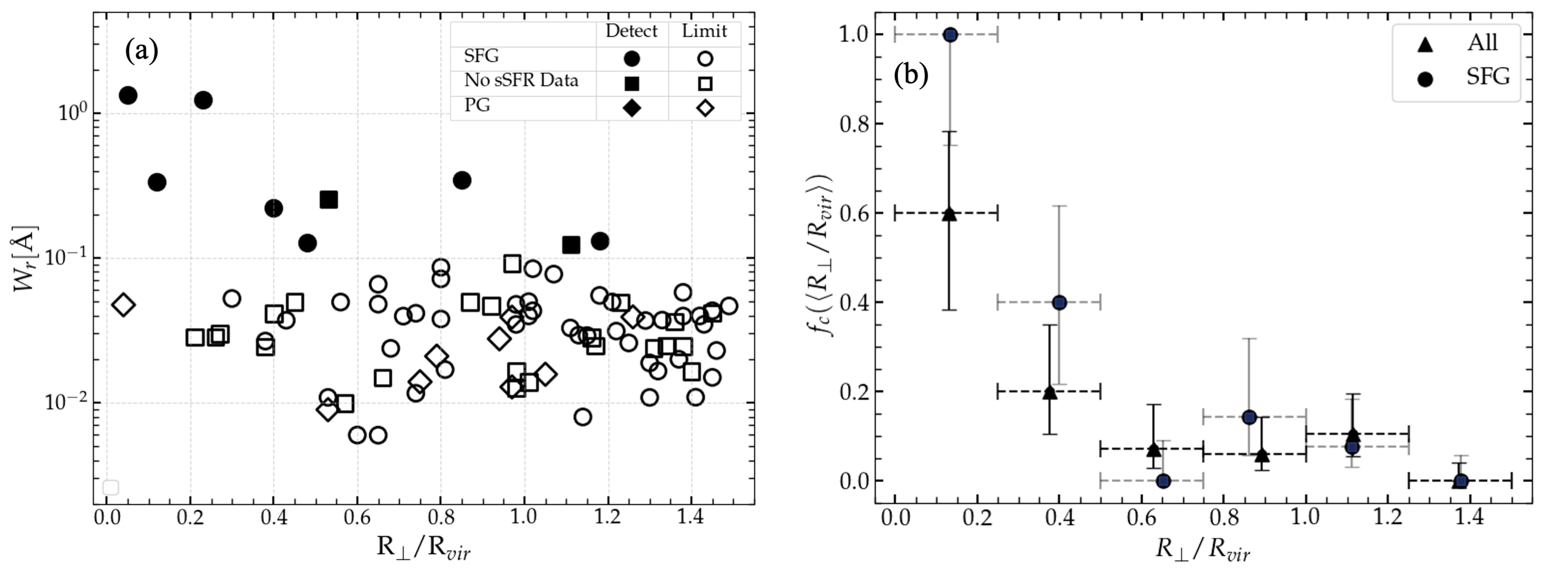}
\caption{
(a) $W_r$ versus $R_\perp/R_{\rm vir}$ for $R_\perp/R_{\rm vir} \leq 1.5$. Solid points are absorbers with $W_r \geq 0.1$~{\AA} and open points are non-absorbers with upper limits $W_{r,{\rm lim}} = 0.1$~{\AA}. (b) The {\CIV} absorption covering fraction, $f_c$, as a function of $R_\perp/R_{\rm vir}$. Triangles represent the full sample, whereas circles represent the SFG subsample. The vertical uncertainties are $1\sigma$ Wilson binomial confidence intervals \citep{wilson27}.  
}
\label{fig:EWvsD}
\end{figure*}
\begin{figure*}[htb]
\includegraphics[width=2.1\columnwidth]{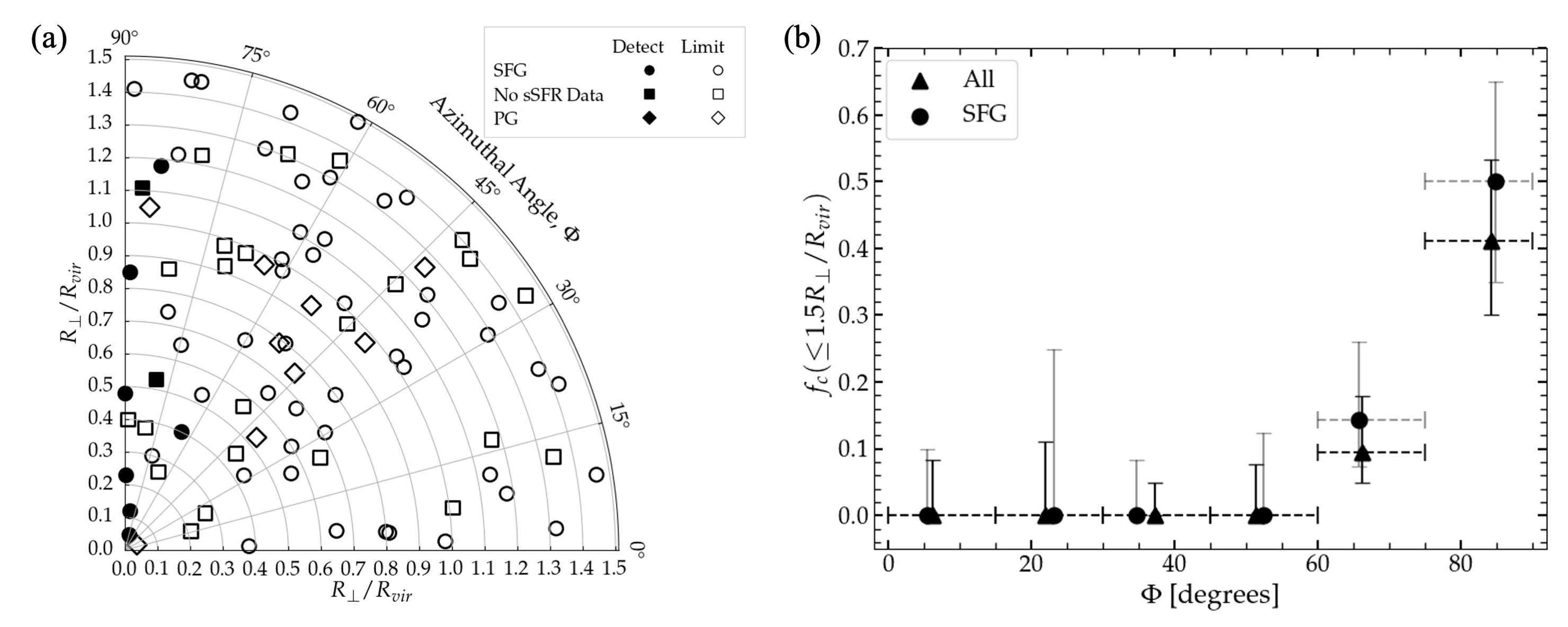}
\vglue -0.1in
\caption{
(a) The azimuthal distribution of galaxy--quasar impact parameters normalized to the virial radius of the host galaxy, ${R_\perp/R_{\rm vir}}$, where ${\Phi = 0^\circ}$ is the projected galaxy major axis and ${\Phi = 90^\circ}$ is the projected galaxy minor axis. Solid points represent absorbing galaxies and open points represent non-absorbing galaxies. Circular points are SFGs, diamond points are PGs, and square points have no measured sSFR. 
(b) The covering fraction, $f_c$, of {\CIV} absorbers inside ${R_\perp\leq 1.5 R_{\rm vir}}$ as a function of azimuthal angle, $\Phi$.  Triangles represent the full sample of galaxies, whereas circular points are the subsample of SFGs. 
}  
\label{fig:EWvsAz}
\end{figure*}

\subsection{Physical Extent and Covering Fraction}
\label{sec:EWvsD}

\label{sec:fcCIV}

In Figure~\ref{fig:EWvsD}(a) we present $W_r$ as a function of $R_\perp/R_{\rm vir}$. Though we searched for {\CIV} absorption as far out as ${R_\perp = 1.5 R_{\rm vir}}$, we found no absorption in the 23 sightlines at $R_\perp \geq 1.2 R_{\rm vir}$. The data are consistent with other works that show a decline in the absorption strength with projected separation from the host galaxy \citep[e.g.,][and many others]{magiicat2, tumlinson17}.

In Figure~\ref{fig:EWvsD}(b), we present the {\CIV} absorption covering fraction, $f_c$,  as a function of $R_\perp/R_{\rm vir}$ in bins of $0.25R_{\rm vir}$. The vertical uncertainties are $1\sigma$  Wilson binomial confidence intervals \citep{wilson27}. For $R_\perp \leq 0.5R_{\rm vir}$, the covering fraction is slightly higher for SFGs compared to the full sample. However, the data do not provide a significant indication that {\CIV} absorbing gas is more homogeneous or less patchy around SFGs.

The mean {\CIV} covering fraction inside the virial radius is $f_c \simeq 0.2$. This finding is consistent with the {\CIV} covering fraction for $W_r \geq 0.1$~{\AA} within $R_{\rm vir}$ of isolated galaxies at $z<1.1$ measured with the QSAGE sample \citep{dutta21}. This would imply that the patchiness of {\CIV}-bearing gas above this detection threshold has remained fairly constant over the last ${\sim  8}$~Gyr. This is interesting given that $dN/dX$ for {\CIV} absorbers with $W_r \geq 0.05$~{\AA} has {\it increased\/} by a factor of $\sim\!3.5\times$ from $z=1$ to $z=0$ \citep[see][]{hasan20}.

However, over the range ${0.25 \leq R_\perp/R_{\rm vir} \leq 1.0}$, the covering fractions from this study are a factor of ${\sim\!5}$--$7\times$ lower than those measured by \citet{garza25a} for ${\langle z \rangle \sim 0.3}$ galaxies. Whereas the minimum threshold for the covering fraction here is $W_r = 0.1$~{\AA}, the minimum threshold of \citeauthor{garza25a} is $\log (N_{\hbox{{\scriptsize C}{\tiny IV}}}/{\rm cm}^{-2} ) = 13.5$.  Based on curve of growth analysis and considering a turbulent/bulk component of $\sim 15$~{\kms} \citep{werk16, mizener19, qu22}, we assess that the two thresholds are directly comparable. 
One possibility for the discrepancy is that the {\CIV} covering fraction substantially decreases from ${z \sim 0.3}$ to ${z \sim 0.03}$. However, $dN/dX$ for {\CIV} absorbers with $W_r \geq 0.05$~{\AA} increased by a factor of $\sim\!1.5\times$ from $z=0.3$ to $z=0$ \citep[][]{hasan20}. An increasing path density of {\CIV} over the last 3.5~Gyr is difficult to reconcile with a factor of several decrease in the {\CIV} CGM covering fraction since $z \simeq 0.3$. 

The discrepancy between the covering fractions may be informing us that galaxy selection plays an important role for interpreting global absorber statistics. For this work, the galaxies were selected to be isolated with no {\it a priori\/} knowledge of CGM absorption. \citet{garza25a} selected galaxies previously studied in absorption from an assortment of previously published works.
The local environments of those samples are not fully controlled, and the majority of the \citeauthor{garza25a}\ galaxies were known to have strong {\HI} absorption and/or multiphase metal-line absorption, even if knowledge of {\CIV} absorption was not known in advance for a fair number of the cases. As such, we favor the interpretation that sample selection explains the elevated covering fractions in the \citeauthor{garza25a} sample relative to those obtained in both this work and in the QSAGE study of \citet{dutta21}.

\subsection{Azimuthal Distribution}
\label{sec:therebewinds}


In {Figure~\ref{fig:EWvsAz}(a)}, we show the distribution of $\Phi$ versus $R_\perp/R_{\rm vir}$, where ${\Phi = 0^\circ}$ is the projected galaxy major axis (planar region) and ${\Phi = 90^\circ}$ is the projected galaxy minor axis (polar region). Absorbers exclusively reside within $30^\circ$ of the projected minor axis of SFGs (though two have unknown sSFR.) All but one of the absorbing galaxies is highly inclined, i.e., $i > 45^\circ$.



A 2D two-sample KS test \citep{fasano87} comparing the distribution of  absorbers and non-absorbers on the $\Phi\,$--$\,R_\perp/R_{\rm vir}$ plane yielded a KS statistic of 0.8353 with $P({\rm KS}) = 1.86\times 10^{-5}$ (${\sigma_{\hbox{\tiny KS}} = 4.3}$). We can thus rule out that the $\Phi\,$--$\,R_\perp/R_{\rm vir}$ distribution of absorbers is consistent with that of non-absorbers at the 99.998\% confidence level. Recall that we have demonstrated that the observed distribution of azimuthal angles for this sample is consistent with being drawn from an unbiased distribution. As such, in this sample, the propensity for absorption to be aligned with the projected minor axes of the host galaxies is a statistically-significant finding.

In Figure~\ref{fig:EWvsAz}(b), we present the covering fraction of {\CIV} absorbers with ${W_r \geq 0.1}$~{\AA} inside ${R_\perp\leq 1.5 R_{\rm vir}} $ as a function of $\Phi$ in bins of $\Delta \Phi = 15^\circ$.  Roughly 30--50\% of the sightlines within $15^\circ $ of the projected minor axis exhibit {\CIV} absorption as do 10--20\% of the sightlines within  $ 15^\circ$ to $30^\circ$ of the minor axis. No {\CIV} absorption with ${W_r \geq 0.1}$~{\AA} was found in the azimuthal range ${0^\circ \leq \Phi < 60^\circ}$ inside ${R_\perp/R_{\rm vir} \leq 1.5}$.

In each azimu
thal bin, the covering fraction of the subsample of SFGs is statistically consistent with that of the full sample. Re-binning the two samples in bins of $\Delta \Phi = 30^\circ$, the covering fractions within $30^\circ$ of the projected minor axes are $f_c({\leq\! 1.5R_\perp/R_{\rm vir}}) \simeq 0.34\pm 0.05$ for the full sample and $f_c({\leq\! 1.5R_\perp/R_{\rm vir}}) \simeq 0.50\pm 0.13$ for the subsample of SFGs. These overlap within the $1\sigma$ uncertainties. A high sSFR in the host galaxy may not be a sufficient criterion for polar-axis {\CIV} absorption in and of itself because $\simeq 50$\% of the SFGs with sightlines within $30^\circ$ of the projected minor axis are non-absorbers. This could imply transient lifetimes for {\CIV}-bearing wind material or a substantially patchy distribution of {\CIV}-bearing gas in winds.

\section{Conclusions and Discussion}
\label{sec:wrapitup}

We studied {\CIV} CGM absorption for 88 ${L_B \geq 0.2L^*_B}$ isolated galaxies at ${0.0017\leq z \leq 0.03}$ that reside within ${R_\perp \leq 1.5 R_{\rm vir}}$ of background quasars sightlines. The galaxy sample was constructed blind to all absorption in the spectra of the background quasars and to all other galaxy properties.  Galaxies with ${\log ({\rm sSFR}/{\rm yr}^{-1}) \geq -10.75}$ were classified as star-forming galaxies (SFGs), whereas those below this sSFR were classified as passive galaxies (PGs). Only 9/88 galaxies in the sample give rise to {\CIV} absorption with $W_r \geq 0.1$~{\AA} within $R_\perp/R_{\rm vir} \leq 1.2$ of their host galaxies. Here, we summarize our key findings. 

(1) For the full galaxy sample, the {\CIV} covering fraction for $W{_r \geq 0.1}$~{\AA} absorbers declines with projected distance $R_\perp/R_{\rm vir}$. For  $R_\perp \leq 0.5R_{\rm vir}$, the covering fraction declines from $f_c \simeq 0.6$ to ${f_c \sim 0.2}$. It then is relatively constant at $f_c \sim 0.1$ for ${0.5 < R_\perp/R_{\rm vir} \leq 1.25}$. Beyond $R_\perp = 1.25R_{\rm vir}$, there is no absorption and $f_c$ vanishes.

(2) Inside the virial radius, $f_c$ for {\CIV} measured in this work at $z \leq 0.03$ is fully consistent with the covering fraction measured inside the virial radius at $0.3\leq z \leq 1.1$ for the QSAGE isolated galaxy sample \citep[see][]{dutta21}. However, it is roughly a factor of ${\sim\!5}$--$7\times$ smaller than that reported at ${z \sim 0.3}$ by \citet{garza25a}. This discrepancy may be due to the very different selection criteria of the absorber--galaxy pairs between the two studies and highlights the importance of galaxy samples constructed blind to any CGM absorption and controlled for environment.


(3) All nine of the {\CIV} absorbers in the full galaxy sample reside within an opening angle of $30^\circ$ of their host galaxy projected minor axes. To a confidence level of 99.998\% ($4.3\sigma$), we can rule out that the azimuthal distribution of absorbers is consistent with the unbiased azimuthal distribution of the sample. The covering fractions as a function of azimuthal angle are $f_c \simeq 0.4$ for ${75^\circ \leq \Phi \leq 90^\circ}$, $f_c \simeq 0.1$ for ${60^\circ \leq \Phi \leq 75^\circ}$, and $f_c \simeq 0$ for $\Phi \leq 60^\circ$. 

(4) The covering fraction for the subsample of star-forming galaxies in the range $75^\circ \leq \Phi \leq 90^\circ$ is $f_c \simeq 0.5$. As such, a high specific star formation rate in a host galaxy may not be a universal predictor for the presence of {\CIV} absorption. If minor-axis {\CIV} absorbing gas arises in biconical outflowing stellar winds in these star-forming galaxies, then the winds must either be quite patchy in {\CIV}-bearing gas and/or the lifetimes of {\CIV} absorbing clouds must be such that half the time few to no such clouds are intersected by a quasar sight line.

\subsection{Is there a Selection Bias?}

Our study was double blind. Quasar sightlines were selected blind to both the presence of intervening galaxies or the detection of any quasar absorption lines. Galaxies were selected blind to potentially associated quasar absorption lines and with no regard to galaxy orientation. 
Nevertheless, some degree of bias might be present, since our galaxy-quasar sample draws on existing COS spectra in the MAST archives. These COS/G160M quasar spectra were acquired for a variety of original scientific objectives, some of which may have been motivated by {\it a priori\/} knowledge of intervening galaxy and/or absorption-line properties that are central to our science objectives. 

After the current research was completed, we conducted an exhaustive literature and archival search using MAST, NED, and NASA/ADS with some assistance from Google's Gemini.
For the nine absorbing galaxies in our sample, we address the following two questions: (1) Could our results be driven by the possibility that the quasars behind the sample galaxies, especially those with detected {\CIV} absorption, were targeted with COS/G160M because the intervening galaxies were known to give rise to {\HI}  and/or any CGM metal-line absorption? (2) Is it possible that the orientations (galaxy inclination and galaxy-quasar azimuthal angle) were factors in the selection of the quasar sightlines for COS/G160M observations?

 For 5/9 galaxies (NGC~3173, 3489, 3941, 4253, and 4517), we find ``no bias;'' there is no evidence to the affirmative for questions (1) and (2). For 2/9 galaxies (NGC 0755 and 1097), we find what we would consider ``virtually no bias.'' NGC~0755 was targeted blind to absorption in a large survey of bright quasars with candidate absorbing foreground galaxies; it was later incorporated into COS-Halos and COS-Dwarfs \citep{tumlinson13, bordoloi-cosdwarfs} and the general {\CIV} survey of \citet{burchett16}. NGC~1097 was targeted blind to absorption in a study by \cite{bowen16} and {\CIV} absorption was detected in its inner CGM. 

For the two remaining galaxies (NGC~2611 and 3067), we find a ``possible mild bias.''  NGC~2611 was not singled out, but was included in the large COS-GTO survey and appeared in \citet{stocke13}. The archived COS spectra were then published in a multi-galaxy study by \citet{keeney17} and \citet[][as part of the COS-Burst collaboration]{heckman17}. As part of COS-DIISC, \citet{koplitz25} re-examined the archived COS absorption spectrum, including {\CIV} absorption around NGC 2611. Importantly, the COS spectrum was obtained only as part of a larger COS survey (and we remind the reader we were blind to the aforementioned studies of NGC~2611). NGC~3067 is intervening to the famous and well-studied quasar 3C~232. Decades ago, NGC~3067 was discovered to have optical {\CaII} and {\NaI} CGM absorption \citep{bosky+wal78, stocke91} and later, {\CIV} and {\SiIV} absorption in COS/G160M spectra \citep{keeney05}. NGC~3067 was then studied by the COS-SAGA Survey team in a multi line-of-sight study focused on the connection between the CGM and satellite galaxies \citep{hamanowicz24}.

Overall, it would appear that there is no strong affirmative to the above questions, apart from NCG~3067. We find it unlikely that bias due to targeting of the quasar sight lines due to the {\it a priori\/} knowledge of absorption in the CGM of the intervening galaxies significantly affects our results.  


\subsection{Discussion}


In this sample of 88 galaxies, it is quite remarkable that ${f_c \sim 0.5}$ exclusively within $30^\circ$ of the galaxy projected minor axes and that ${f_c=0}$ as azimuthal angle approaches the projected major axes (${0^\circ \leq \Phi \leq 60^\circ}$). It may be that a strong signal of intermediate-ionization wind material funneled into biconical outflows is a common feature of present-epoch galaxies. Low-ionization minor-axes outflows are commonly found at ${z\sim 0}$ \citep[e.g.,][]{heckman00, martin09, chen.y-m10, bae18}. \citet{concas19} observed that low-ionization and neutral winds were entrained from the galactic disk and powered by stellar feedback, whereas highly-ionized outflows originated from the accretion disk powered by AGN feedback.

In simulations, \citet{nelson19} found that the degree of collimation in biconical winds increases with cosmic time, being a fairly universal phenomenon at ${z < 1}$. They suggested that this redshift evolution is coupled to a rise of orderly rotating gas structures that contribute to ``sculpting'' outflow collimation. \citet{peroux20} found that, as redshift decreases below ${z \sim 1}$ to the present epoch, the metallicity is increasingly enhanced along the galaxy projected minor axes of galaxies relative to their major axes. No such signature is found at higher redshift and \citeauthor{peroux20} argue that early CGM metallicity mixing requires a few Gyr to distill out based on its spatial-kinematic origin.

The combined effects of higher metallicity winds with higher degrees of collimation along galaxy projected minor axes at the present epoch may contribute to an explanation for our findings. To explore this possibility, we compared the observed {\CIV} absorption velocities to idealized CGM geometries and kinematics. Employing the spatial-kinematic absorption models (SKAM) developed by \citet{skamI, skamII}, we explored simple biconical outflows \citep[e.g.,][]{fielding20, fielding22} and extended planar accretion \citep[][]{stewart11, hafen22, trapp22, stern24} as constrained by each galaxy's inclination, quasar impact parameter, and the galaxy-quasar azimuthal angle.  As detailed in \citet{mcroom-thesis}, we adopted a fiducial model with a wind opening angle of $\Theta_w = 30^\circ$, a wind base radius of $\rho_{w,0} = 10$~kpc, and a constant wind velocity of $V_w =150$~{\kms}. We also assumed a planar accretion opening angle of $\Theta_a = 10^\circ$, an accretion disk interface of $\rho_{a,0} = 25$~kpc, and a spiraling infall with circular and axial components $(V_c, V_\rho) =  (220,200)$~{\kms} \citep[see Figs.~5, 11, and 14 of][for schematics of the geometries and kinematics]{skamI}. 

The upshot of the SKAM modeling was that for 7/9 galaxies, SKAM indicates the gas is consistent with wind-driven outflows, and for 2/9 galaxies SKAM can be made consistent with either wind and/or accretion CGM spatial kinematics. Taken at face value, we find that winds are heavily favored for most of the absorbing galaxies, but that accretion could explain some of the relative galaxy-absorber kinematics. Of particular interest is that the azimuth angles are robustly found within $30^\circ$ of the project minor axes for all nine galaxy-absorber pairs, yet SKAM indicates that wind-driven flows may not be the sole origin of the absorption in every case.  However, if {\CIV}-bearing gas is common in planar accreting material, we would expect that such accretion structures would be observed in quasar sightlines spanning the full range of azimuthal angles, especially major-axis orientations ($\Phi \sim 0^\circ$). Yet, this is not observed. At least in this sample of galaxies, we conclude that {\CIV} absorption from an accretion component of the baryon cycle is very infrequent.

For there to be no detected absorption along the major axes and intermediate azimuthal angles, we would require physical mechanisms that do not favor the creation of {\CIV} absorbing clouds in these regions of the CGM.  In the multi-component multiphase ionization models of \citet{sameer24}, 11/47 galaxies at ${0.11 \leq z \leq 0.88}$ exhibit {\CIV} absorption, of which 8/11 are aligned within $30^\circ$ of the projected minor axes, whereas 3/11 lie in the range $\Phi \leq 60^\circ$; these three reside at ${z>0.3}$, a look-back time exceeding ${\simeq 3.5}$~Gyr. To best explain {\CIV} absorption, \citeauthor{sameer24} ruled out collisional ionization, favoring time-dependent photoionization cloud models that are more highly ionized and warmer than photoionization equilibrium models. Using this same absorber--galaxy sample, \citet{kacprzak26} found that clouds not consistent with disk rotation (winds?) were more highly ionized and had ${\approx\!0.5}$~dex higher metallicity. Nearer to the minor-axis, clouds have lower $N({\HI})$, lower hydrogen density, and an enhanced non-thermal line broadening compared to major-axis clouds.

Comparing the intermediate-ionization {\CIV} in this work, which shows only minor-axis absorption, to the observed spatial-kinematic distributions of low-ionization {\MgII} and high-ionization {\OVI}, we note that both are very different than what we found for {\CIV}. Both {\MgII} and {\OVI} show a similar bimodality in their covering fractions (enhanced at both the major and minor axes), whereas {\CIV} exhibits absolutely zero propensity to live outside a bipolar ``cone'' geometry with opening angle $\pm 30^\circ$.  {\CIV}-bearing CGM gas may isolate the interface between these low- and high-ionization regimes; it may comprise intermediate-ionization clouds that have been enriched and lifted out of the disk by feedback that are observed to be mixing and cooling via non-equilibrium photoionization within bipolar outflow cones.

\section*{ACKNOWLEDGMENTS}
We thank the anonymous referee, whose careful reading and thoughful comments improved this manuscript.
Support for program AR-14772 (PI Wakker) was provided by STScI, operated by ARAA, Inc.\ (NASA grant NAS5-26555). This research used 
(1) the HyperLEDA (univ-lyon1.fr) and NED (NASA/IPAC) databases,
(2) the Siena Galaxy Atlas (DOE High Energy Physics grant DE-SC0020086 and NSF grant AST-1616414), and 
(3) the Heraklion Extragalactic Catalogue (v2), \citep{2023hell.confE...1K} hosted by FORTH.

\bibliographystyle{aasjournal}  
\bibliography{main}

\end{document}